\documentclass[preprint,pra,aps]{revtex4-1}

\usepackage{amsmath,esint}
\usepackage{amsfonts}
\usepackage{amssymb}
\usepackage{gensymb}
\usepackage{siunitx}
\usepackage{verbatim}
\usepackage{graphicx}
\usepackage{times}
\usepackage{xfrac}

\begin{document}

\title{Photonic Dirac monopoles and skyrmions: spin-1 quantization}

\author{Todd Van Mechelen}
\author{Zubin Jacob}
\email{zjacob@purdue.edu}
\affiliation{Birck Nanotechnology Center and Purdue Quantum Center, Department of Electrical and Computer Engineering, Purdue University, West Lafayette, Indiana 47907, USA}

\begin{abstract}
We introduce the concept of a photonic Dirac monopole, appropriate for photonic crystals, metamaterials and 2D materials, by utilizing the Dirac-Maxwell correspondence.  We start by exploring vacuum where the reciprocal momentum space of both Maxwell's equations and the massless Dirac equation (Weyl equation) possess a magnetic monopole. The critical distinction is the nature of magnetic monopole charges, which are integer valued for photons but half-integer for electrons. This inherent difference is directly tied to the spin and ultimately connects to the bosonic or fermionic behavior. We also show the presence of photonic Dirac strings, which are line singularities in the underlying Berry gauge potential. While the results in vacuum are intuitively expected, our central result is the application of this topological Dirac-Maxwell correspondence to 2D photonic (bosonic) materials, as opposed to conventional electronic (fermionic) materials. Intriguingly, within dispersive matter, the presence of photonic Dirac monopoles is captured by nonlocal quantum Hall conductivity - i.e. a spatiotemporally dispersive gyroelectric constant. For both 2D photonic and electronic media, the nontrivial topological phases emerge in the context of massive particles with broken time-reversal symmetry. However, the bulk dynamics of these bosonic and fermionic Chern insulators are characterized by spin-1 and spin-\sfrac{1}{2} skyrmions in momentum space, which have fundamentally different interpretations. This is exemplified by their contrasting spin-1 and spin-\sfrac{1}{2} helically quantized edge states. Our work sheds light on the recently proposed quantum gyroelectric phase of matter and the essential role of photon spin quantization in topological bosonic phases. 
\end{abstract}

\maketitle

\section{Introduction}

Dirac's pioneering paper \cite{Dirac60} showed that if magnetic monopoles are found in nature, their magnetic charges $Q$ would be quantized in units of the elementary charge $e$ of the electron,
\begin{equation}
2\frac{eQ}{h}\in\mathbb{Z}.
\end{equation}
$h$ being the Planck constant. This is the earliest example of topological quantization - fundamentally different from second quantization arising in quantum field theories. Although there exists no experimental proof of magnetic monopoles \cite{Preskill1984} to date, there is ample evidence of quantized topological charges in reciprocal (energy-momentum) space. Specifically, the appearance of such monopoles in the band structure of solids indicates the presence of quantized topological invariants, like the Chern number \cite{Thouless1982} and $\mathbb{Z}_2$ index \cite{Kane_2_2005}. Ultimately, experimental observables such as the quantum Hall conductivity can be traced back to the existence of this quantized topological charge \cite{Xu613,Laughlin1981}.

There have been significant efforts to construct synthetic gauge potentials that mimic these monopole physics in cold atoms \cite{Zoller2005} and spin ice \cite{Morris411}. One striking example is the realization of non-Abelian gauge theories with Yang-Lee monopoles \cite{sugawa_observation_2016}. The topological field theory of light has surfaced in knotted solutions of Maxwell's equations \cite{Kedia2013,Stone2016}, as well as the uncertainty relations for photons \cite{BB2012}. Along side this, there have been important recent developments to formulate topological properties for photons utilizing photonic crystals and metamaterials \cite{Lu2014,Rechtsman2013,Khanikaev2013,CTChan2016,Wang2016,Silveirinha2015,Haldane2008,Haldane2008_2,HafeziM.2013,ShanuiFan2017,Shuang2015,GLYBOVSKI20161,Papasimakis:10}. The pioneering work in topological photonic crystals has shown the existence of edge states robust to disorder. In the previously explored scenarios, the photonic crystal unit cell is carefully structured to obtain an additional degree of freedom (artificial gauge field) - quite often realized on a graphene-like honeycomb lattice. This approach was first implemented by Haldane for spinless (scalar) electrons in his seminal paper on the parity anomaly \cite{Haldane1988}. However, it remains an open question whether robust topological photonic edge states can occur in atomic matter. The role of photon spin and its quantization is yet another unresolved problem since previous theories have focused exclusively on pseudo-spin-\sfrac{1}{2} phenomena \cite{Lu622,Chen2016,Yangeaaq1221,WangLuyang2016}.

Our spin-1 theory \cite{Hu2018} is fundamentally different in this respect since we do not ignore the polarization (spin) state of the photon, which cannot be neglected for a real gauge (vector) field. In our case, the topological theory is manifestly bosonic as it is connected to the winding of the gauge field itself - not pseudo-spin degrees of freedom. Another fundamental aspect of our theory is the inclusion of dispersion within matter, i.e. frequency and momentum dependence of conductivity, such that topological invariants emerge naturally from the global behavior of optical constants. For example, it has been shown that nonlocal gyrotropic \cite{VanMechelen2018} and magnetoelectric \cite{van_mechelen_2017} media will host massless spin-1 quantized edge states with massive-like photons in the bulk. Thus, it is necessary to understand the concept of bosonic Dirac monopoles and the influence of integer spin in topological photonic phases of matter.

In this paper, we elucidate the fundamental difference between the magnetic monopoles appearing in Maxwell's equations and the Dirac equation. Our work shows that a magnetic monopole appears for both photons and massless fermions in the reciprocal energy-momentum space - even for vacuum. Using a Dirac-Maxwell correspondence, we identify the bosonic and fermionic nature of magnetic monopole charge, which is inherently present in the relativistic theories of both particles. While the results in vacuum are expected, we apply this topological theory to 2D photonic (bosonic) materials, in contrast to conventional electronic (fermionic) materials. The specific 2D photonic materials considered in this paper are gyroelectric which possess antisymmetric components of the conductivity tensor. We exploit the Dirac-Maxwell correspondence to show how dispersive gyroelectric media can support topologically massive particles, which are interpreted as photonic skyrmions. However, the differences in spin between bosons and fermions alter the behavior of these bulk skyrmions as well as their corresponding Chern numbers. We then analyze the unique topological edge states associated with nontrivial spin-1 and spin-\sfrac{1}{2} skyrmions, which exhibit opposing helical quantization. This clearly shows how the integer and half-integer nature of monopoles is ultimately tied to the differing bosonic and fermionic spin symmetries. Our work sheds light on the recently proposed quantum gyroelectric phase of matter \cite{VanMechelen2018} which supports unidirectional transverse electro-magnetic (TEM) edge states with open boundary conditions (vanishing fields at the edge) - unlike any known phase of matter till date.

In the context of geometric phases, the concept of magnetic charges has a rich history starting from the pioneering works of Pancharatam, Berry, Chiao and Wu \cite{chruscinski_geometric_2004}. Unification of these geometric phases for bosons and fermions was shown for massive 3D particles using a relativistic quantum field theory \cite{BB1987}. In this paper, our focus is massless 3D particles and topologically massive 2D particles \cite{Shi2018,Mortensen2018,Horsley2018}, as well as the direct demonstration of gauge discontinuities in Maxwell's and Weyl's equations. Our derivation does not utilize quantum field theoretic techniques and appeals only to the spin representation of the two particles. We note that spin quantization is fundamentally different from topological charges encountered in real space for OAM beams \cite{barnett_natures_2016,Gawhary2018}, polarization singularities \cite{Doug2014} and polarization vortices \cite{Shabanov2008}. This is due to the central concept of gauge discontinuity in magnetic monopole quantization, which is related to the topological field theory of bosons and fermions. We function in momentum space of Maxwell's equations as opposed to real space so our work is specifically suited to develop topological invariants in the band structure of photonic crystals and wave dispersion within metamaterials \cite{Sihvola2005,Li2018}. One important application of our current technique is in uncovering unique electromagnetic phases of matter displaying the quantum gyroelectric effect (QGEE) \cite{VanMechelen2018}. Our unified perspective also sheds light on recent developments of quantized bosonic Hall conductivity \cite{Lu2012,Senthil2013,Metlitski2013,Vishwanath2013} and topological bosonic phases of matter \cite{Wen2011,Chen1604}, as opposed to fermionic phases \cite{Kane2010}.

Skyrmions have a storied past in condensed matter - appearing in both real and momentum space of topological systems. In real space, these localized topological defects were first discovered in chiral magnets and quantum Hall ferromagnets but have also been observed in Bose-Einstein condensates and superconductors \cite{han_skyrmions_2017}. The behavior of these magnetic skyrmions is intimately tied to the Dzyaloshinskii-Moriya (DM) interaction \cite{Nagaosa2013} which generates the nontrivial winding of the spin structure. In momentum space, skyrmions often characterize the monopoles arising in the band structure of solids and are emergent phenomena in topological insulators and superconductors \cite{bernevig2013topological}. By contrast, photonic skyrmions are a very recent field of interest. A classical optical analog of skyrmion-like behavior has been reported using surface plasmon polaritons \cite{Tsesseseaau0227}. This work focuses on photonic skyrmions in momentum space which ultimately govern the equations of motion of a topological electromagnetic field. The physics of these topological fields manifest in nontrivial windings of a spin-1 vector as opposed to a spin-\sfrac{1}{2} vector.

\textbf{Note:} For clarity, all 3D vectors will be denoted with a vector arrow $\vec{A}=(A_x,~A_y,~A_z)$, while we reserve boldface for 2D vectors $\mathbf{A}=(A_x,~A_y)$. The manuscript theme is the Dirac-Maxwell correspondence which directly compares bosonic and fermionic topological field theories. Throughout, the subscript $s=1$ stands for spin-1 photons and the subscript $s=\sfrac{1}{2}$ denotes spin-\sfrac{1}{2} electrons.

\section{Three dimensions: massless particles}

\subsection{Dirac-Maxwell correspondence}

The correspondence between Dirac's and Maxwell's equations is best expressed in the Riemann-Silberstein (R-S) basis \cite{Bialynicki-Birula2013,Barnett2014}, which utilizes a vector wave function for light. Using this representation, we develop a topological field theory of the vacuum photon. In the R-S basis, we combine the electric $\vec{E}$ and magnetic $\vec{H}$ fields into a complex superposition,
\begin{equation}
\vec{\Psi}=\frac{1}{\sqrt{2}}(\vec{E}+i\vec{H}),
\end{equation}
where $i=\sqrt{-1}$ is the imaginary unit and the electromagnetic fields are associated with plane waves. We strongly emphasize that relativity requires vectorial representations for spin-1 bosonic fields and spinor-\sfrac{1}{2} representations for fermionic fields. Spin-0 particles constitute scalar fields while spin-2 particles, such as gravitons, are described by tensor fields. Therefore, to unravel the topological bosonic properties of light, we cannot work in a restricted subspace ignoring components of the electromagnetic field. Simultaneously, we do not describe polarizations separately. In the R-S basis, Maxwell's equations in vacuum can be combined into a first-order wave problem as follows,
\begin{equation}\label{eq:Maxwell}
i\vec{k}\times\vec{\Psi} =H_1\vec{\Psi}=\omega\vec{\Psi},
\end{equation}
which we label as spin $s=1$. Here, $\omega$ is the frequency of light and we consider dynamical fields over all frequencies and wave vectors, not simple static fields. We can thus unambiguously identify a Hamiltonian for light,
\begin{equation}\label{eq:Spin1}
H_1(\vec{k})=\vec{k}\cdot\vec{S}=k_xS_x+k_yS_y+k_zS_z.
\end{equation}
$\vec{k}=(k_x,~k_y,~k_z)$ is the momentum of the plane wave in vacuum and $\vec{S}=(S_x,~S_y,~S_z)$ are the set of SO(3) antisymmetric matrices,
\begin{equation}\label{eq:SO3}
S_x=\begin{bmatrix}
0 & 0 & 0 \\
0 & 0 & -i \\
0 & i & 0 \\
\end{bmatrix}, \qquad S_y=\begin{bmatrix}
0 & 0 & i \\
0 & 0 & 0 \\
-i & 0 & 0 \\
\end{bmatrix}, \qquad S_z=\begin{bmatrix}
0 & -i & 0 \\
i & 0 & 0 \\
0 & 0 & 0 \\
\end{bmatrix}.
\end{equation}
These operators obey the familiar Lie algebra $[S_i,S_j]=i\epsilon_{ijk}S_k$ which encode information about integer spin. Notice our photonic Hamiltonian $H_1=\vec{k}\cdot\vec{S}$ represents optical helicity, i.e. the projection of spin $\vec{S}$ along the direction of momentum $\vec{k}$. This is further clarified on direct comparison with massless Dirac fermions (Weyl fermions), which are the supersymmetric partners of the massless photon \cite{Stephen2001}. The Weyl equation is expressed as,
\begin{equation}\label{eq:Dirac}
H_{\sfrac{1}{2}}\psi=E\psi, 
\end{equation}
where the massless Dirac Hamiltonian $H_{\sfrac{1}{2}}$, corresponding to spin $s=\sfrac{1}{2}$, is identified with electronic helicity,
\begin{equation}\label{eq:Spin1half}
H_{\sfrac{1}{2}}(\vec{k})=\vec{k}\cdot\vec{\sigma}=k_x\sigma_x+k_y\sigma_y+k_z\sigma_z.
\end{equation}
$\vec{\sigma}=(\sigma_x,~\sigma_y,~\sigma_z)$ are the Pauli matrices of SU(2) and obey the identical Lie algebra $[\sigma_i,\sigma_j]=2i\epsilon_{ijk}\sigma_k$,
\begin{equation}\label{eq:SU2}
\sigma_x=\begin{bmatrix}
0 & 1\\
1 & 0
\end{bmatrix}, \qquad \sigma_y=\begin{bmatrix}
0 & -i\\
i & 0
\end{bmatrix}, \qquad \sigma_z=\begin{bmatrix}
1 & 0\\
0 & -1
\end{bmatrix}.
\end{equation}
Both particles are massless and satisfy an analogous helicity equation. However, the critical difference is revealed in the group operations of the particular particle; encapsulated by the SO(3) antisymmetric matrices for the spin-1 photon [Eq.~(\ref{eq:SO3})] and the SU(2) Pauli matrices for the spin-\sfrac{1}{2} Weyl fermion [Eq.~(\ref{eq:SU2})].

\subsection{Helical eigenstates}

\begin{figure}
\centering
\includegraphics[width=\linewidth]{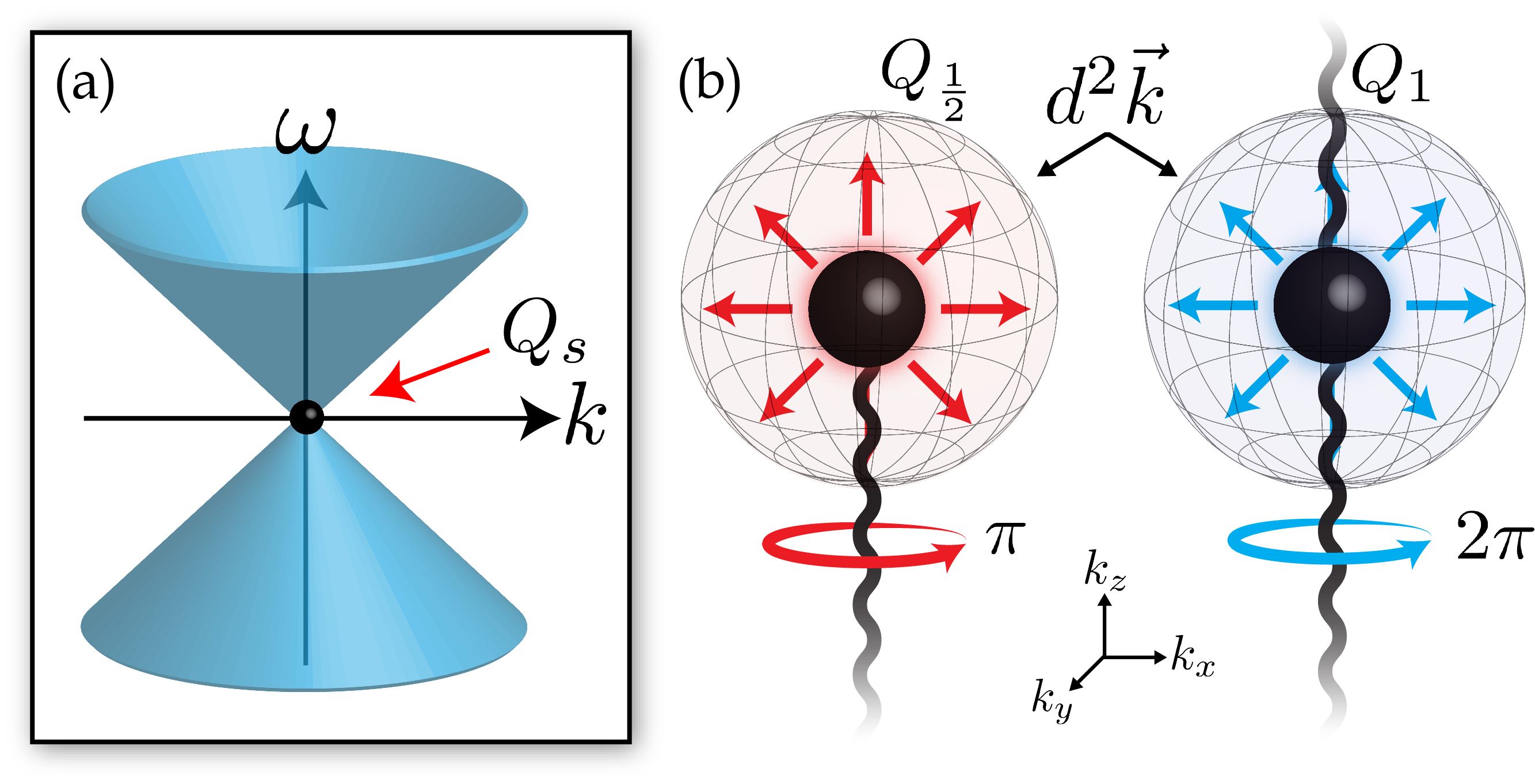}
\caption{(a) Linear dispersion (light cone) of the 3D massless photon and electron $\omega_\pm=E_\pm=\pm k$. At the origin of the momentum space $\vec{k}=0$ sits a magnetic monopole with quantized charge. This singularity is often called a Weyl point and is quantized to the spin of the particle $Q_s=s$. Integer and half-integer spin quantization is connected to bosonic and fermionic statistics respectively. (b) Dirac monopoles (Berry curvature) $\vec{F}_s=\vec{\nabla}_k\times\vec{A}_s$ of the massless electron $Q_{\sfrac{1}{2}}=\sfrac{1}{2}$ and photon $Q_1=1$ in momentum space. The monopole charge acts as a source for the magnetic field $\vec{\nabla}_{k}\cdot\vec{F}_s=4\pi Q_s\delta^3(\vec{k})$ and arises due to the discontinuous behavior in the spin eigenstates. Notice that the flux through any surface enclosing the monopole is necessarily quantized $Q_s=(4\pi)^{-1}\oiint \vec{F}_s\cdot d^2\vec{k}$. This monopole is accompanied by a string of singularities in the underlying gauge potential $\vec{A}_s$. Any closed path around the equator of the string produces a quantized Berry phase  $\gamma_{s}=\oint\vec{A}_s\cdot d\vec{k}= 2\pi Q_s$. The accumulated phase in $\vec{k}$-space is fundamentally tied to the spin of the particle $\mathcal{R}_s(2\pi)=\exp(i\gamma_s)=(-1)^{2s}$.}
\label{fig:LinearDispersion}
\end{figure}

We now solve for the eigenstates of the above Hamiltonians. As expected, Maxwell and Weyls' equations possess two helical degrees of freedom. For the photon [Eq.~(\ref{eq:Spin1})], these are conventional right- and left-handed circular polarization,
\begin{equation}\label{eq:Circular}
H_1\vec{e}_\pm=\pm k ~\vec{e}_\pm, \qquad \vec{e}_\pm(\vec{k})=\frac{1}{\sqrt{2}}(\hat{\theta}\pm i\hat{\phi}),
\end{equation}
where $\theta$ and $\phi$ are the spherical polar coordinates of $\vec{k}$ and $k=|\vec{k}|$ is the magnitude of the wave vector. The photon is massless and therefore linearly dispersing in vacuum $\omega_\pm =\pm k$. Similarly, the eigenstates of the Weyl equation [Eq.~(\ref{eq:Spin1half})] are comprised of two massless helical spinors, which are represented as,
\begin{equation}\label{eq:Spinor}
H_{\sfrac{1}{2}}\psi_\pm = \pm k~\psi_\pm, \qquad \psi_+(\vec{k})=\begin{bmatrix}
\cos (\theta/2) \\
\sin (\theta/2)e^{i\phi}
\end{bmatrix},\qquad \psi_-(\vec{k})=\begin{bmatrix}
\sin (\theta/2) \\
-\cos (\theta/2)e^{i\phi}
\end{bmatrix}.
\end{equation}
Indeed, these states are also linearly dispersing $E_\pm=\pm k$. An important observation can be made in $\vec{e}_\pm$ and $\psi_\pm$. The eigenstates are ill-defined at the origin of the momentum space $\vec{k}=0$, since they are arbitrarily dependent on $\theta$ and $\phi$ at this point. In fact, by parameterizing $\theta$ as the inclination from $k_z$, the eigenstates are not well-behaved at the north $\theta=0$ or south  $\theta=\pi$ poles either - they are multivalued at both points. Such discontinuous behavior is impossible to remove and results from choosing a particular gauge for the eigenstates. This is the underlying source for Dirac monopoles and strings. The linear dispersion (light cone) of the massless helical states is displayed in Fig.~\ref{fig:LinearDispersion}(a).

\subsection{Spin quantization in photonic Dirac monopoles and strings}

In vacuum $\vec{k}$-space, we discover a magnetic Dirac monopole for both Maxwell's and Weyl's equations but with intrinsic differences. This is demonstrated by first defining the magnetic flux in momentum space - i.e. the Berry curvature. For the photon, the Berry curvature of either right- or left-handed helicity can be found from the circular eigenstates derived in Eq.~(\ref{eq:Circular}),
\begin{equation}
\vec{F}_1^\pm=-i\vec{\nabla}_{k}\times[{\vec{e}_\pm}^*\cdot(\vec{\nabla}_{k}~\vec{e}_\pm)].
\end{equation}
For the massless electron, the analogous Berry curvature is found by evaluating the spinor eigenstates in Eq.~(\ref{eq:Spinor}), 
\begin{equation}
\vec{F}_{\sfrac{1}{2}}^\pm=-i\vec{\nabla}_{k}\times[\psi_\pm^\dagger\vec{\nabla}_{k}~\psi_\pm].
\end{equation}
Here, $\vec{\nabla}_{k}=\sum_j \hat{j}~\partial_{k_j}$ is the gradient operator in 3D momentum space. Note that the Berry curvature is a vector in three dimensions but a scalar in two dimensions. On evaluating the Berry curvature for both particles with positive and negative helicities ($\pm$), we find that $\vec{F}_s^\pm=\pm \vec{F}_s$ possesses a Dirac monopole, 
\begin{equation}\label{eq:DiracMonopole}
 \vec{F}_s= Q_s\vec{F}.
\end{equation}
$\vec{F}$ being the magnetic field of a Dirac monopole in $\vec{k}$-space,
\begin{equation}
\vec{F}(\vec{k})=\frac{\vec{k}}{k^3}.
\end{equation}
Note that $Q_s$ in Eq.~(\ref{eq:DiracMonopole}) is the topological magnetic charge which generates the magnetic field. This quantity is fundamentally different for the two particles,
\begin{equation}
Q_s= s.
\end{equation}
$s$ is precisely the spin of the particle, which takes integer $s=1$ or half-integer $s=\sfrac{1}{2}$ values for bosons or fermions respectively. We emphasize that the magnetic monopole charge is naturally quantized, 
\begin{equation}
Q_s=\frac{1}{4\pi}\oiint \vec{F}_s\cdot d^2\vec{k}.
\end{equation}
The charge is located at the origin $\vec{k}=0$ of the momentum space, exactly where the eigenstates are ill-defined, and acts as a source for the magnetic field $\vec{\nabla}_{k}\cdot\vec{F}_s=4\pi Q_s\delta^3(\vec{k})$. Notice that the magnetic monopole charge of the photon,
\begin{equation}
Q_1=2Q_{\sfrac{1}{2}}=1,
\end{equation}
is exactly twice the electron due to integer spin. The monopole charge for each helicity has opposite signs $Q_s^\pm=\pm Q_s$. This ensures the net charge vanishes $Q^+_s+Q^-_s=0$ at the origin $\vec{k}=0$; as expected due to time-reversal symmetry in vacuum \cite{Young2012}. A visualization of the magnetic flux is shown in Fig.~\ref{fig:LinearDispersion}(b).

We note that the photonic Dirac monopole is accompanied by a string of singularities in the underlying gauge potential. This Dirac string is unobservable as it is a gauge dependent phenomenon but sheds light on the fundamental differences between electrons and photons. The Berry gauge potential for the massless photon and electron can be evaluated using the eigenstates in Eq.~(\ref{eq:Circular}) and (\ref{eq:Spinor}) respectively,
\begin{equation}
\vec{A}_1^\pm=-i{\vec{e}_\pm}^*\cdot(\vec{\nabla}_{k}~\vec{e}_\pm), \qquad \vec{A}_{\sfrac{1}{2}}^\pm=-i\psi_\pm^\dagger\vec{\nabla}_{k}~\psi_\pm.
\end{equation}
Upon solving for $\vec{A}_s^\pm =\pm \vec{A}_s$, we again find a clear dependence on the magnetic monopole charge $Q_s$ which is different for bosons and fermions,
\begin{equation}\label{eq:DiracString}
\vec{A}_s(\vec{k})=Q_s\frac{1-\cos\theta}{k\sin\theta}\hat{\phi},
\end{equation}
and $\vec{F}_s=\vec{\nabla}_{k}\times\vec{A}_s$ reproduces the Berry curvature in Eq.~(\ref{eq:DiracMonopole}). The gauge potential is singular along the $k_z$-axis, at $\theta=0$ and $\pi$, where the eigenstates are multivalued. This line singularity that originates at the monopole and extends to infinity is known as a Dirac string. Fig.~\ref{fig:LinearDispersion}(b) displays a visualization of the Dirac monopole and strings for both massless particles. We note that the above equations are traditionally found in the theory of magnetic charges in real space \cite{Preskill1984} - not momentum space. Following this, quantization of magnetic charge naturally emerges from the requirement of a single-valued wave function in the presence of singular (multivalued) gauge potentials. Our rigorous derivation is unique as it unifies the momentum space of Maxwell's equations and the Weyl equation. This makes it ideally suited for extension to topological theories of band structure in photonic crystals and wave dispersion in metamaterials.

\subsection{Berry phase}

We now provide a detailed comparison of $\vec{k}$-space Pancharatnam-Berry phase (hereon called geometric phase) for photons and electrons, that arises from their corresponding spin properties. The geometric phase calculated for any closed path on the $\vec{k}$-sphere is gauge invariant,
\begin{equation}
\gamma_s=\oint \vec{A}_s\cdot d\vec{k}=\iint \vec{F}_s\cdot d^2\vec{k}.
\end{equation}
$\gamma_s$ is the geometric phase and is equivalent to the flux of Berry curvature $\vec{F}_s$ through a surface bounded by the path. In this case, we see that $\iint \vec{F}_s\cdot d^2\vec{k}=Q_s\iint d\Omega$ is exactly the solid angle $\Omega(C)$ traced along the $\vec{k}$-sphere,
\begin{equation}
\gamma_s=Q_s\Omega(C),
\end{equation}
where $C$ designates the bounded path. We now consider a closed path around a great circle of the $\vec{k}$-sphere (eg: the equatorial path $k_z=0$), which encloses the monopole. For massless particles, this is equivalent to rotating the fields back into themselves. The accumulated phase must be quantized,
\begin{equation}\label{eq:GeometricPhase}
\gamma_s=2\pi Q_s.
\end{equation}
This is the momentum space manifestation of Dirac's quantization condition $2Q_s\in \mathbb{Z}$ which ensures the massless particles acquire the same phase under a $2\pi$ or $-2\pi$ rotation. We clearly see that geometric phases in $\vec{k}$-space are dependent on the spin of the particle,
\begin{equation}
\exp(i\gamma_s)=(-1)^{2Q_s}.
\end{equation}
Notice that $\exp\left(i\gamma_{\sfrac{1}{2}}\right)=-1$ and $\exp\left(i\gamma_{1}\right)=+1$ are antisymmetric or symmetric under a $\pm 2\pi$ rotation depending on the spin $Q_s=s$. Ultimately, the geometric phase of $\gamma_{\sfrac{1}{2}}=\pi$ or $\gamma_1=2\pi$ is tied to the fermionic or bosonic statistics of the particle. We note that this geometric phase $\gamma_{\sfrac{1}{2}}=\pi$  is routinely encountered for massless Dirac fermions in graphene \cite{Mazrzari2011,Novoselov2006}. However, the direct correspondence with spin-1 massless photons $\gamma_1=2\pi$ has not been pointed out to date. Our results suggest that a thin wire supporting Dirac fermions would yield Chiao-Tomita phases \cite{Chiao1986} exactly half the value of photons. We also note that spin-momentum locking is a universal property in photonics \cite{VanMechelen2016,kalhor2016universal,Pendharker:18} which arises entirely from the transversality $\vec{k}\cdot\vec{\Psi}=0$ of electromagnetic waves in vacuum. This phenomenon can be explained with causal boundary conditions on evanescent fields and does not necessarily require topological considerations \cite{Bliokh1448}. For example, conventional surface plasmon polaritons (SPPs) and waveguide modes show spin-momentum locking but these are not related to any topologically protected edge states or nontrivial phases.

\subsection{Rotational symmetries}

The nuance behind integer and half-integer geometric phases [Eq.~(\ref{eq:GeometricPhase})] is explained more rigorously by considering the operations of the rotational (spin) groups. Maxwell's equations [Eq.~(\ref{eq:Maxwell})] transform under the SO(3) group $\mathcal{R}_1(\alpha)=\exp\left(i\alpha\hat{n}\cdot\vec{S}\right)$, where $\alpha$ is the angle subtended about an axis $\hat{n}$. This is true for all vector fields. Conversely, the Weyl equation [Eq.~(\ref{eq:Dirac})] transforms under the SU(2) group $\mathcal{R}_{\sfrac{1}{2}}(\alpha)=\exp\left(i\alpha\hat{n}\cdot\vec{\sigma}/2\right)$, characteristic of spinors. Although SO(3) and SU(2) obey the same Lie algebra, the group representations are inequivalent. The distinction is evident under a cyclic rotation,
\begin{equation}
\mathcal{R}_s(2\pi)=(-1)^{2s}.
\end{equation}
Notice that the accumulated phase is different depending on the particle species. This is due to the fact that fermions are antisymmetric $\mathcal{R}_{\sfrac{1}{2}}(2\pi)=-1$ under rotations, while bosons are symmetric $\mathcal{R}_1(2\pi)=+1$ and this behavior is guaranteed by the spin-statistics theorem \cite{Pauli1940}. The difference fundamentally changes the interpretation of fermionic and bosonic topologies \cite{Regnault_2013}.

\section{Two dimensions: topologically massive particles}

\subsection{Dirac-Maxwell correspondence}

Up to this point, we have only considered the 3D dynamics of the vacuum photon and its analogies with the Weyl fermion. Now we shift to the 2D domain to harness these topological properties and elucidate the fundamental role of spin in nontrivial phases of matter. Nontrivial 2D materials are characterized by an integer topological invariant - the Chern number $C\in\mathbb{Z}$. In electronics, these materials are often called Chern insulators \cite{Jotzu2014} because they are insulating in the bulk but host metallic one-way edge states that are robust to disorder. In the long wavelength limit $k\approx 0$, the simplest fermionic Chern insulator is described by the 2D Dirac equation \cite{bernevig2013topological},
\begin{equation}\label{eq:FermionChern}
\mathcal{H}_{\sfrac{1}{2}}\psi=E\psi, \qquad \mathcal{H}_{\sfrac{1}{2}}(\mathbf{k})=v(k_x\sigma_x+k_y\sigma_y)+\Lambda(k)\sigma_z.
\end{equation}
Equation~(\ref{eq:FermionChern}) is essentially identical to the Weyl equation [Eq.~(\ref{eq:Spin1half})] except we have replaced the $z$-component of the momentum with a Dirac mass $k_z\to\Lambda(k)$. We have also introduced the Fermi velocity $v$ to characterize the effective speed of electrons within the material. It is easy to check that $\Lambda(k)$ breaks time-reversal symmetry but preserves rotational symmetry about the $z$-axis, $\mathcal{R}_{\sfrac{1}{2}}(\alpha)=\exp\left(i\alpha\sigma_z/2\right)$. The meat of the topological physics lies in this spatially dispersive Dirac mass \cite{shen2011topological},
\begin{equation}
\Lambda(k)=\Lambda_0-\Lambda_2k^2.
\end{equation}
$\Lambda_0=\Lambda(0)$ opens a band gap and $\Lambda_2$ accounts for the curvature of the energy bands. Importantly, when $\Lambda_0\Lambda_2>0$ there is so-called \textit{band inversion} \cite{Zhu2012} and the effective mass changes sign within the dispersion $\Lambda(k_i)=0$, precisely at $k_i=\sqrt{\Lambda_0/\Lambda_2}$. The quadratic momentum dependence $k^2=\mathbf{k}\cdot\mathbf{k}$ is also crucial to regularize the long wavelength theory \cite{VanMechelen2018,Ryu2010}. This means we can project the planar momentum space onto the surface of the Riemann sphere $\mathbb{R}^2\simeq S^2$, a necessary constraint for continuum topological field theories.

We now study the equivalent 2D dynamics of the photon - the bosonic Chern insulator. As anticipated, the 2D Maxwell theory is the supersymmetric partner of the 2D Dirac theory \cite{VanMechelen2018,Dunne1999} and takes an analogous form,
\begin{equation}\label{eq:BosonChern}
\mathcal{H}_{1}\vec{\Psi}=\omega\vec{\Psi}, \qquad \mathcal{H}_{1}(\mathbf{k})=v(k_xS_x+k_yS_y)+\Lambda(k)S_z.
\end{equation}
Equation~(\ref{eq:BosonChern}) is formally equivalent to the 3D Maxwell equation [Eq.~(\ref{eq:Spin1})] with the substitution of a mass term $k_z\to\Lambda(k)$. Here, $v=1/\sqrt{\varepsilon}$ is the effective speed of light which is governed by the dielectric permittivity $\varepsilon>1$. Like the Dirac equation [Eq.~(\ref{eq:FermionChern})], time-reversal symmetry is broken but rotational symmetry is preserved about the $z$-axis, $\mathcal{R}_1(\alpha)=\exp\left(i\alpha S_z\right)$. There is one caveat however; the photonic wave function $\vec{\Psi}$ is slightly altered since we only retain transverse-magnetic (TM) waves in two dimensions,
\begin{equation}
\vec{\Psi}=\frac{1}{\sqrt{2}}\left(\sqrt{\varepsilon}E_x, ~\sqrt{\varepsilon}E_y, ~i H_z\right)=\frac{1}{\sqrt{2}}\left(\sqrt{\varepsilon}\mathbf{E},~ i H_z\right).
\end{equation}
The transverse-electric (TE) component cannot couple to a 2D material as all electrical currents lie in the $x$-$y$ plane. Nevertheless, the underlying topological physics remain unchanged.
\subsection{Dispersive transverse conductivity}
Our central result is that the above mentioned Maxwell Hamiltonian can possess a mass term arising from dispersion of optical constants. Still, one might question the seemingly ad hoc insertion of a photonic mass $\Lambda(k)$ for two reasons: 1. \textit{Are Maxwell's equations still gauge invariant?} 2. \textit{Does this mass have any physical origin?} The answer is \textit{yes} to both \cite{Dunne1999,BOYANOVSKY1986483}. In fact, it is nothing but the Hall conductivity \cite{Hall1879},
\begin{equation}
\varepsilon\Lambda(k)=\sigma_H(k)=\sigma_0-\sigma_2k^2.
\end{equation}
Remarkably, our result shows that the Hall conductivity for 2D photons plays the exact same role as the Dirac mass for 2D electrons. We note that the Hall conductivity is related to the anti-symmetric components of the conductivity tensor.  $\sigma_0=\sigma_H(0)$ is the conventional static (DC) component which opens a band gap in the vacuum dispersion. This property of low energy bandgap is fundamentally similar to the role of the Dirac mass for fermions.  $\sigma_2$ is the nonlocal (momentum dependent) correction to $\sigma_H$ which dictates the curvature of the photonic bands. Until very recently, the momentum dependence of $\sigma_H$ had never been considered for topological purposes \cite{VanMechelen2018}. This type of behavior can also be generalized to its high-frequency (AC) equivalent in the context of nonlocal gyrotropy, but we restrict ourselves to the low-energy limit $\omega\approx 0$ for simplicity. In this limit, nonlocal Hall conductivity defines the quantum gyroelectric phase of matter.

\begin{figure}
\centering
\begin{minipage}[c]{0.55\linewidth}
\includegraphics[width=\linewidth]{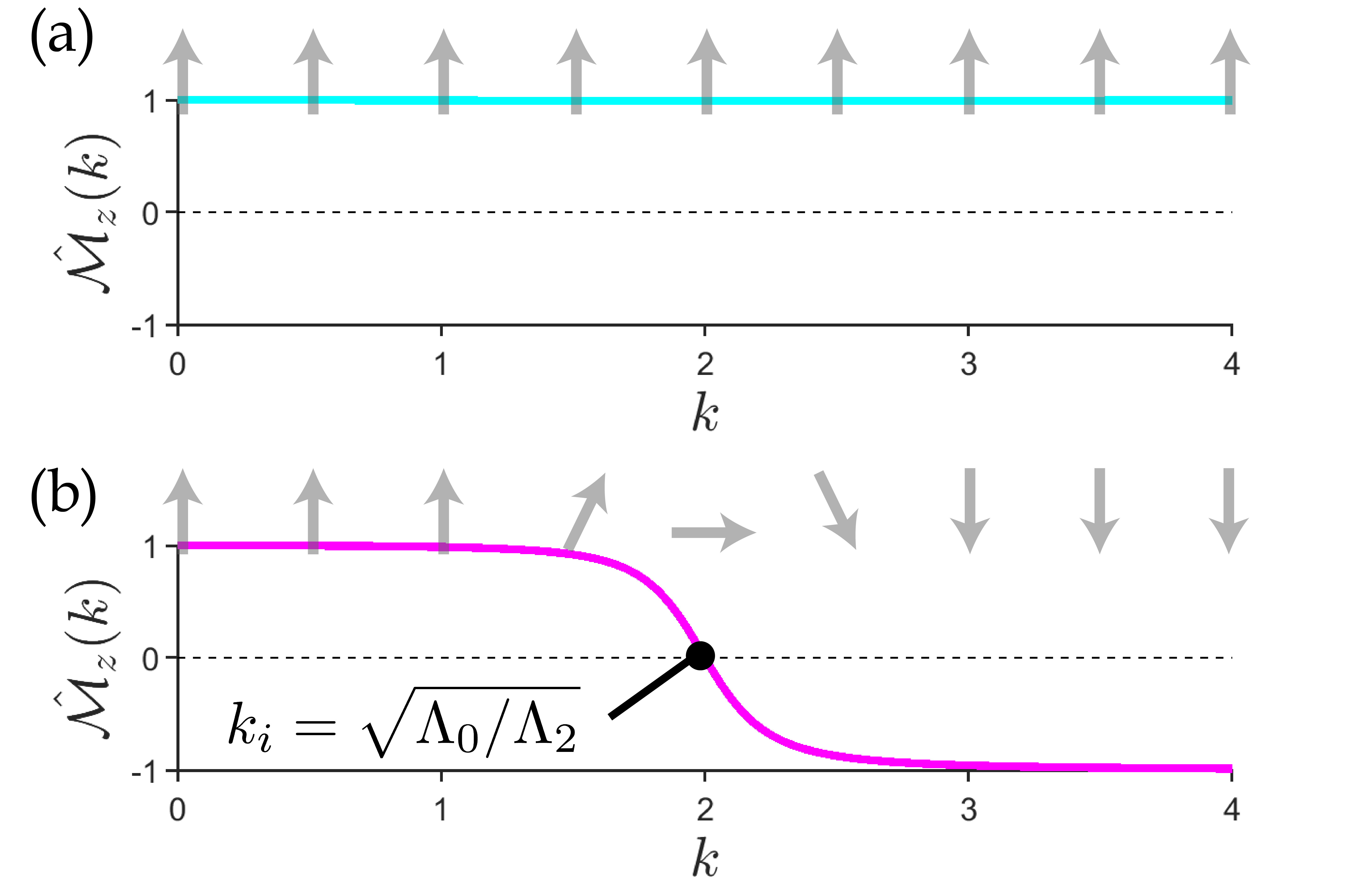}
\end{minipage}\hfill
\begin{minipage}[c]{0.45\linewidth}
\caption{Spin expectation value $\hat{\mathcal{M}}_z(k)$ as a function of $k$. (a) $N=0$ skyrmion with no band inversion $\Lambda_0\Lambda_2<0$. The spin returns to initial state $\hat{\mathcal{M}}_z(0)=\hat{\mathcal{M}}_z(\infty)$ and the total winding is trivial. (b) $N=1$ skyrmion with band inversion $\Lambda_0\Lambda_2>0$. In this case, the spin flips direction $\hat{\mathcal{M}}_z(0)\neq\hat{\mathcal{M}}_z(\infty)$ and the total winding is nontrivial. $k_i$ labels the band inversion point where $\hat{\mathcal{M}}_z(k_i)=0$ passes through zero. This point must occur for the spin to flip directions and can only be removed at a topological phase transition.}
\label{fig:spin_texture}
\end{minipage}
\end{figure}

\subsection{Spin-1 photonic skyrmions}

The electronic [Eq.~(\ref{eq:FermionChern})] and photonic [Eq.~(\ref{eq:BosonChern})] Hamiltonians can be written in a more suggestive form by introducing the skyrmion spin vector $\vec{\mathcal{M}}=(\mathcal{M}_x,~\mathcal{M}_y,~\mathcal{M}_z)$,
\begin{equation}\label{eq:SkyrmionVector}
\mathcal{H}_1=\vec{\mathcal{M}}\cdot\vec{S}, \qquad \mathcal{H}_{\sfrac{1}{2}}=\vec{\mathcal{M}}\cdot\vec{\sigma}.
\end{equation}
As we can see, this new vector $\vec{\mathcal{M}}$ has replaced the original 3D wave vector $\vec{k}$ in the massless equations and closely resembles the Zeeman interaction \cite{han_skyrmions_2017}. Indeed, the spin precesses about an axis formed by $\vec{\mathcal{M}}$,
\begin{equation}
\dot{\vec{S}}=-i[\vec{S},\mathcal{H}_1]=g_1\vec{\mathcal{M}}\times\vec{S}.
\end{equation}
It is important to reiterate that $\vec{S}$ represents spin-1 operators while $\vec{\sigma}$ is spin-\sfrac{1}{2}. This is exemplified by the fact that bosonic (vector) particles possess gyromagnetic $g$-factors of $g_{1}=(Q_1)^{-1}=1$, while fermionic (spinor) particles have $g$-factors of $g_{\sfrac{1}{2}}=(Q_{\sfrac{1}{2}})^{-1}=2$, 
\begin{equation}
\dot{\vec{\sigma}}=-i[\vec{\sigma},\mathcal{H}_{\sfrac{1}{2}}]=g_{\sfrac{1}{2}}\vec{\mathcal{M}}\times\vec{\sigma}.
\end{equation}
The Larmor frequency $\Omega_s$ is fundamentally different between the two. The skyrmions precess at different rates depending on the spin representation,
\begin{equation}
\Omega_s=g_s\mathcal{M}, \qquad g_s=(Q_s)^{-1},
\end{equation}
where $\mathcal{M}=|\vec{\mathcal{M}}|$ is the magnitude of the skyrmion vector.

Note though, the skyrmion vector $\vec{\mathcal{M}}(\mathbf{k})$ is a function of a 2D momentum $\mathbf{k}$ and actually describes a \textit{parametric surface} $\vec{\mathcal{M}}(\mathbf{k})=\left(vk_x,~vk_y,~\Lambda(k)\right)=\left(v\mathbf{k},~\Lambda(k)\right)$. The eigenstates assume an identical form with the substitution of $\vec{k}\to\vec{\mathcal{M}}$,
\begin{equation}\label{eq:Skyrmion}
\mathcal{H}_1\vec{e}_\pm = \pm \mathcal{M}~\vec{e}_\pm, \qquad \mathcal{H}_{\sfrac{1}{2}}\psi_\pm = \pm \mathcal{M}~\psi_\pm.
\end{equation}
$\vec{e}_\pm$ are the right- and left-handed helical eigenstates derived in Eq.~(\ref{eq:Circular}) and $\psi_\pm$ are the equivalent spinors in Eq.~(\ref{eq:Spinor}). The dispersion relation for each of the eigenstates reads,
\begin{equation}
\omega_\pm(k)=E_\pm(k)=\pm \mathcal{M}(k)=\pm\sqrt{v^2k^2+\Lambda^2(k)},
\end{equation}
which are clearly gapped since $\mathcal{M}(0)=|\Lambda_0|$. These states have acquired mass in 2D. The critical difference of these new eigenstates is that the polar coordinate $\theta$ no longer parametrizes the inclination from $k_z$. Instead, it is governed by the spatially dispersive mass $\tan\theta(k)=vk/\Lambda(k)$, which is a function of the \textit{in-plane} momentum $\mathbf{k}$. We can understand this phenomenon more clearly by evaluating the spin expectation value along the $\hat{z}$ direction,
\begin{equation}
\langle S_z\rangle_\pm={\vec{e}_\pm}^*\cdot S_z\vec{e}_\pm=\pm Q_1\hat{\mathcal{M}}_z,
\end{equation}
where $\hat{\mathcal{M}}_z(k)=\Lambda(k)/\mathcal{M}(k)=\cos\theta(k)$ is a normalized vector. Notice the spin comes in units of bosonic charge $Q_1=1$, as we would expect for an integer particle $s=1$. Analogously, the half-integer skyrmion $s=\sfrac{1}{2}$ arises in units of fermionic charge $Q_{\sfrac{1}{2}}=\sfrac{1}{2}$, 
\begin{equation}
\left\langle \sigma_z/2\right\rangle_\pm=\psi^\dagger_\pm (\sigma_z/2)\psi_\pm=\pm Q_{\sfrac{1}{2}}\hat{\mathcal{M}}_z.
\end{equation}
At $k=0$, the spin points directly along $\hat{\mathcal{M}}_z(0)=\textrm{sgn}[\Lambda_0]$. However, as the momentum increases, $\vec{\mathcal{M}}$ tilts away from the $z$-axis and in some cases can flip directions entirely $\hat{\mathcal{M}}_z(\infty)=-\textrm{sgn}[\Lambda_2]$. This is a nontrivial topology. A depiction of trivial and nontrivial $\hat{\mathcal{M}}_z(k)$ as a function of $k$ is presented in Fig.~\ref{fig:spin_texture}.

\begin{figure}
\centering
\includegraphics[width=\linewidth]{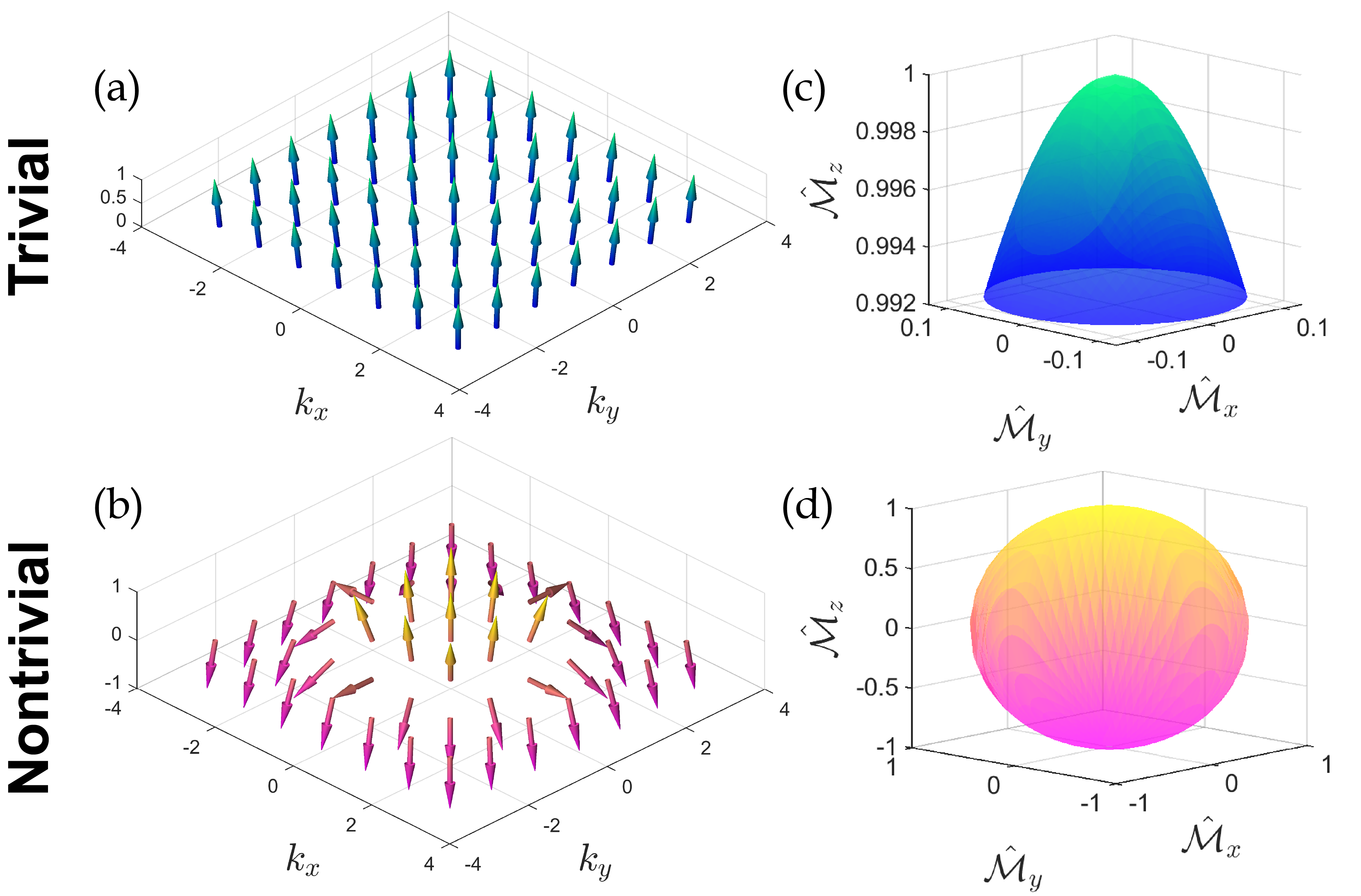}
\caption{Left: spin texture $\hat{\mathcal{M}}(\mathbf{k})$ as a function of $\mathbf{k}$ for trivial and nontrivial skyrmions. (a) $N=0$ skyrmion with no band inversion $\Lambda_0\Lambda_2<0$. As an example, we have let $v=0.5$, $\Lambda_0=4$ and $\Lambda_2=-2$. Since the spin returns to initial state within the dispersion $\hat{\mathcal{M}}_z(0)=\hat{\mathcal{M}}_z(\infty)$, the total winding is trivial. (b) $N=1$ skyrmion with band inversion $\Lambda_0\Lambda_2>0$. To demonstrate, we have let $v=0.5$, $\Lambda_0=4$ and $\Lambda_2=2$. In this case, the spin flips direction within the dispersion $\hat{\mathcal{M}}_z(0)\neq\hat{\mathcal{M}}_z(\infty)$ and the total winding is nontrivial. Right: spin texture $\hat{\mathcal{M}}$ of the skyrmion projected on the unit sphere. As the momentum varies over all possible values, $\hat{\mathcal{M}}(\mathbf{k})$ can perform either a (c) retracted or (d) full evolution over the unit sphere. This corresponds to a total solid angle of $\Omega=0$ or $4\pi$ respectively.}
\label{fig:spin}
\end{figure}

\subsubsection*{\textbf{An aside: the zero helicity (longitudinal) state}}

For completeness, there is technically one additional eigenstate associated with the photonic Hamiltonian [Eq.~(\ref{eq:BosonChern})] - the zero helicity (longitudinal) state,
\begin{equation}
\mathcal{H}_1\vec{e}_0=0, \qquad \vec{e}_0=\hat{\mathcal{M}}=\vec{\mathcal{M}}/\mathcal{M}.
\end{equation}
$\vec{e}_0$ is a completely flat band $\omega_0=0$ and represents the electrostatic limit (irrotational fields). This band belongs to the Hilbert space but can be removed from the spectrum by enforcing $\mathbf{k}\cdot\mathbf{D}=0$ at zero frequency, which implies there is no static charge present. Moreover, since $\vec{e}_0={\vec{e}_0}^*$ can always be chosen real, the Chern number of this band necessarily vanishes $C_0=0$.

\subsection{Skyrmion magnetic field}

We are now ready to assess the Berry curvature. In two dimensions, the Berry curvature is a scalar and characterizes the ``magnetic'' flux through the planar momentum space $\mathbb{R}^2$. Since our long wavelength theory is regularized, this is equivalent to the flux through the Riemann sphere $S^2$. For the 2D photon, the Berry curvature $\mathcal{F}_1^\pm$ is found by varying the in-plane momentum $\mathbf{k}$ of the right- and left-handed eigenstates $\vec{e}_\pm$,
\begin{equation}
\mathcal{F}^\pm_1=-i(\partial_{k_x}{\vec{e}_\pm}^*\cdot\partial_{k_y}\vec{e}_\pm-\partial_{k_y}{\vec{e}_\pm}^*\cdot\partial_{k_x}\vec{e}_\pm).
\end{equation}
The Berry curvature $\mathcal{F}_{\sfrac{1}{2}}^\pm$ of the 2D electron $\psi_\pm$ is derived in a similar fashion,
\begin{equation}
\mathcal{F}^\pm_{\sfrac{1}{2}}=-i(\partial_{k_x}\psi^\dagger_\pm\partial_{k_y}\psi_\pm-\partial_{k_y}\psi^\dagger_\pm\partial_{k_x}\psi_\pm).
\end{equation}
Just like the 3D massless particles [Eq.~(\ref{eq:DiracMonopole})], the Berry curvature $\mathcal{F}^\pm_{s}=\pm \mathcal{F}_s$ comes in units of quantized magnetic charge $Q_s=s$,
\begin{equation}
\mathcal{F}_{s}=Q_s\mathcal{F}.
\end{equation}
This emergent magnetic field $\mathcal{F}$ is generated by the momentum dependent variations in the spin texture $\hat{\mathcal{M}}=\vec{\mathcal{M}}/\mathcal{M}$,
\begin{equation}
\mathcal{F}=\hat{\mathcal{M}}\cdot(\partial_{k_x}\hat{\mathcal{M}}\times\partial_{k_y}\hat{\mathcal{M}})=\vec{F}\cdot d^2 \vec{\mathcal{M}}.
\end{equation}
$\mathcal{F}$ is precisely the magnetic field of a skyrmion \cite{Nagaosa2013} and has several profound interpretations. Mathematically, its the Jacobian and dictates the degree of continuous mapping from the momentum space (the Riemann sphere) onto the unit sphere $\hat{\mathcal{M}}$, i.e. $S^2\to S^2$. In another context, it tells us the differential flux of the Dirac monopole $\vec{F}$ onto the parametric surface $\vec{\mathcal{M}}(\mathbf{k})$,
\begin{equation}
\vec{F}=\frac{\vec{\mathcal{M}}}{\mathcal{M}^3}, \qquad d^2 \vec{\mathcal{M}}=\partial_{k_x}\vec{\mathcal{M}}\times\partial_{k_y}\vec{\mathcal{M}},
\end{equation}
where $d^2\vec{\mathcal{M}}$ is the surface normal. As the momentum varies over all possible values, the spin vector $\vec{\mathcal{M}}(\mathbf{k})$ can enclose the monopole any number of times. \textit{Hence, the total magnetic flux counts the number of $Q_s$ monopoles enclosed by the skyrmion spin vector $\vec{\mathcal{M}}$},
\begin{equation}
N=\frac{1}{4\pi}\iint_{\mathbb{R}^2}\mathcal{F}dk_xdk_y=\frac{1}{4\pi}\iint_{\mathbb{R}^2}\hat{\mathcal{M}}\cdot(\partial_{k_x}\hat{\mathcal{M}}\times\partial_{k_y}\hat{\mathcal{M}})dk_xdk_y, \qquad N\in \mathbb{Z}.
\end{equation}
This is known as the skyrmion (or winding) number. Since the momentum space is bounded on the Riemann sphere $\mathbb{R}^2\simeq S^2$, the skyrmion number $N$ is guaranteed to be an integer. A visualization of the $\hat{\mathcal{M}}$ unit sphere for trivial and nontrivial skyrmions is displayed in Fig.~\ref{fig:spin}.

\subsection{Chern insulators}

The Chern number $C_s$ is directly proportional to the skyrmion number $N$ but has a very different meaning depending on the particle species. It counts twice the total magnetic charge of the skyrmion,
\begin{equation}
C_s=\frac{1}{2\pi}\iint_{\mathbb{R}^2}\mathcal{F}_sdk_xdk_y=\frac{Q_s}{2\pi}\iint_{\mathbb{R}^2}\mathcal{F}dk_xdk_y=2Q_sN.
\end{equation}
For spin-\sfrac{1}{2} skyrmions, the Chern number is an integer $C_{\sfrac{1}{2}}=N\in \mathbb{Z}$ and is indistinguishable from the skyrmion number itself. Spin-1 skyrmions are quite different by comparison; the Chern number is an \textit{even} integer $C_1=2N\in 2\mathbb{Z}$. From fermionic Chern arguments, one would expect to always find an even number of photonic edge states - but this is not the case \cite{Lu2012,Senthil2013,Metlitski2013,Vishwanath2013}. Although a widely held belief, the conventional bulk-boundary correspondence fails for spin-1 bosons \cite{Tiwari2018}. We will demonstrate this fact explicitly.

\begin{figure}
\centering
\includegraphics[width=\linewidth]{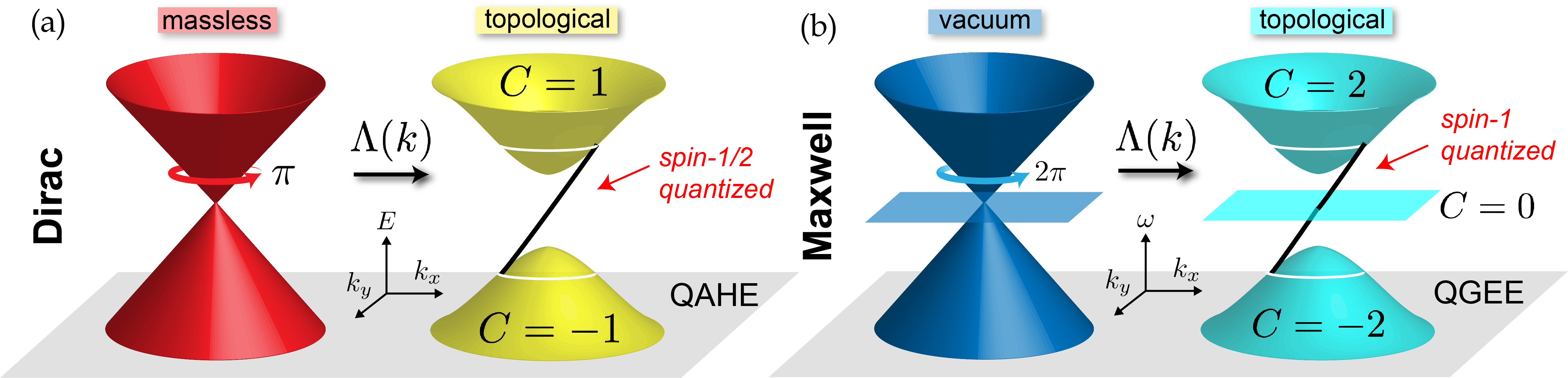}
\caption{Dispersion relation of the bulk and gapless edge bands (black lines) of the topologically massive 2D particles. (a) The conventional fermionic Chern insulator is characterized by a spin-\sfrac{1}{2} skyrmion (Dirac equation). (b) The bosonic Chern insulator is described by a spin-1 skyrmion (Maxwell's equations). The bulk Chern number $C_s=2Q_sN$ depends on both the magnetic charge (spin) $Q_s=s$ and the skyrmion number $N\in\mathbb{Z}$. This corresponds to integer phases for electrons $C_{\sfrac{1}{2}}\in\mathbb{Z}$ but \textit{even} integer phases for photons $C_1\in 2\mathbb{Z}$. At low energy, a band gap is formed at $E=\omega=0$ by a spatially dispersive effective mass $\Lambda(k)=\Lambda_0-\Lambda_2k^2$. (a) For the 2D electron, this is simply the Dirac mass. (b) For the 2D photon, this mass is equivalent to a nonlocal Hall conductivity $\varepsilon\Lambda(k)=\sigma_H(k)$. In the presence of band inversion $\Lambda_0\Lambda_2>0$, there is a point where the effective mass changes sign $\Lambda(k_i)=0$, precisely at $k_i=\sqrt{\Lambda_0/\Lambda_2}$. The massless helically quantized edge states touch the bulk bands at this point. This is known as the quantum anomalous Hall effect (QAHE) for electrons and the quantum gyroelectric effect (QGEE) for photons \cite{VanMechelen2018}. The flat longitudinal band $\omega_0=0$ is shown for completeness and represents the electrostatic limit (irrotational fields). However, this band can be removed from the spectrum by requiring that all static charges vanish.}
\label{fig:dispersion}
\end{figure}

Utilizing our spin vector $\vec{\mathcal{M}}$ defined in Eq.~(\ref{eq:SkyrmionVector}), the skyrmion magnetic field $\mathcal{F}$ in circular polar coordinates $\mathbf{k}=k(\cos\phi,~\sin\phi)$ reads,
\begin{equation}
\mathcal{F}(k)=\frac{vk[\Lambda(k)-vk\Lambda'(k)]}{[v^2k^2+\Lambda^2(k)]^{3/2}}=\sin\theta(k)\partial_k\theta(k)=-\partial_k\hat{\mathcal{M}}_z(k).
\end{equation}
Due to rotational symmetry, $\mathcal{F}(k)$ depends only on the magnitude of $k$. The geometric interpretation is clear - it describes variations in the solid angle $\mathcal{F}(k)dk d\phi=d\Omega(\mathbf{k})$ traced by $\hat{\mathcal{M}}(\mathbf{k})$. Integrating the magnetic flux over all momenta, we acquire the skyrmion number,
\begin{equation}
N=\frac{1}{2}[\cos\theta(0)-\cos\theta(\infty)]=\frac{1}{2}[\hat{\mathcal{M}}_z(0)-\hat{\mathcal{M}}_z(\infty)]=\frac{1}{2}\left(\mathrm{sgn}[\Lambda_0]+\mathrm{sgn}[\Lambda_2]\right).
\end{equation}
When band inversion is present $\Lambda_0\Lambda_2>0$, the $z$-component of the spin vector $\hat{\mathcal{M}}_z(0)\neq\hat{\mathcal{M}}_z(\infty)$ flips directions within the dispersion. This represents north $\theta=0$ and south $\theta=\pi$ poles on the unit sphere, which means $\hat{\mathcal{M}}(\mathbf{k})$ traces out a full solid angle, regardless of the relative magnitudes of $\Lambda_0$ and $\Lambda_2$. This is equivalent to saying the parametric surface $\vec{\mathcal{M}}(\mathbf{k})$ always encloses a monopole $N=\pm 1$. In the trivial regime $\Lambda_0\Lambda_2<0$, the $z$-component $\hat{\mathcal{M}}_z(0)=\hat{\mathcal{M}}_z(\infty)$ returns to its initial state at either the north or south poles and $\vec{\mathcal{M}}(\mathbf{k})$ never encloses a monopole $N=0$. Consequently, the Chern number in the nontrivial phase equates to $C_s=\pm 2Q_s$, which is an integer for the electron $C_{\sfrac{1}{2}}=\pm 1$, but an even integer for the photon $C_1=\pm 2$. In conventional spin-\sfrac{1}{2} and pseudo-spin-\sfrac{1}{2} problems, a large Chern number $|C_{\sfrac{1}{2}}|>1$ corresponds to multiple gapless edge states within the bulk topological band gap. This is not true for spin-1 bosonic particles. For $|C_1|=2$ there is a \textit{single} spin-1 quantized edge state within the topological band gap which is illustrated in Fig.~(\ref{fig:dispersion}).

\subsection{Topological edge states}\label{subsec:Edge}

We now solve for the topology protected edge states of both particles. We stress that for both spin-1 and spin-\sfrac{1}{2} phases, there is exactly one unidirectional solution at the edge. This makes intuitive sense because a single monopole $N=\pm 1$ exists in the band structure. A nontrivial skyrmion $N=\pm 1$ corresponds to either a forward or backward propagating edge state - forward for $N=+1$ and backward for $N=-1$. We take the boundary in the $x$ dimension such that $k_y$ is still a good quantum number. We then look for solutions of the form $\vec{\Psi}^\textrm{e}_\pm(x,y)=\vec{\Psi}^\textrm{e}_\pm(x)e^{ik_y y}$ and $\psi^\textrm{e}_\pm(x,y)=\psi^\textrm{e}_\pm(x)e^{ik_y y}$ that satisfy the boundary condition at infinity $\vec{\Psi}^\textrm{e}_\pm(x=+\infty)=\psi^\textrm{e}_\pm(x=+\infty)=0$. We also impose topological \textit{open boundary conditions} \cite{Hatsugai1993,avila_topological_2013} at the interface,
\begin{equation}
\vec{\Psi}^\textrm{e}_\pm(x=0^+)=\psi^\textrm{e}_\pm(x=0^+)=0.
\end{equation}
If this constraint is satisfied simultaneously, a solution will exist at any interface (even vacuum) because the edge state is insensitive to fields in the $x<0$ region.

Substituting into the Hamiltonians [Eq.~(\ref{eq:FermionChern}) and (\ref{eq:BosonChern})] and applying boundary conditions, the topological edge states emerge. For photonic spin-1 states we have,
\begin{equation}
\vec{\Psi}_\pm^\textrm{e}(x)=\begin{bmatrix}
\sqrt{\varepsilon}E_x\\ \sqrt{\varepsilon}E_y\\ i H_z
\end{bmatrix}^\textrm{e}_\pm=\Psi_0\begin{bmatrix}
1\\ 0\\\mp i
\end{bmatrix}\left(e^{-\eta_1x}-e^{-\eta_2x}\right).
\end{equation}
Carrying out the same procedure, the electronic spin-\sfrac{1}{2} states are expressed as,
\begin{equation}
\psi_\pm^\textrm{e}(x)=\psi_0\begin{bmatrix}
1 \\ \pm i
\end{bmatrix}\left(e^{-\eta_1x}-e^{-\eta_2x}\right).
\end{equation}
The wave functions of the spin-1 and spin-\sfrac{1}{2} particles appear quite similar. The fundamental difference lies in the fact that $\vec{\Psi}$ is a vector (bosonic) field and its polarization state is defined in \textit{real space}. $\psi$ is a spinor (fermionic) field - its polarization state is more abstract as it lives in a complex space. Notice there are two characteristic decay scales for the edge states $\eta_{1,2}$, like a damped harmonic oscillator, but in spatial frequency. These are the quadratic roots of the secular equation,
\begin{equation}
\Lambda_0+\Lambda_2(\eta^2-k_y^2) \mp v\eta=0.
\end{equation}
If $\textrm{sgn}[\Lambda_0]=\textrm{sgn}[\Lambda_2]=+1$ the skyrmion number is $N=+1$ and only a forward propagating solution $(+)$ exists. On the other hand, if $\textrm{sgn}[\Lambda_0]=\textrm{sgn}[\Lambda_2]=-1$ the skyrmion is $N=-1$ and only the backward propagating solution $(-)$ is permitted. $\eta_{1,2}$ characterize the degree of confinement at a particular momentum $k_y$ and are solved straightforwardly,
\begin{equation}
\eta_{1,2}(k_y) =\frac{1}{2|\Lambda_2|}\left[v\pm\sqrt{v^2+4\Lambda_2(\Lambda_2k_y^2-\Lambda_0)}\right].
\end{equation}
The spatial width of the wave packet depends on the size of the band gap formed by $\Lambda_0$ and $\Lambda_2$. However, regardless of their relative magnitudes, as long as $\Lambda_0\Lambda_2>0$ a solution always exists within the band gap - they are topologically protected.

Intriguingly, the edge waves are also \textit{helically quantized} along the direction of propagation $\hat{k}=\hat{y}$,
\begin{equation}
S_y\vec{\Psi}_\pm^\textrm{e}=\pm Q_1\vec{\Psi}_\pm^\textrm{e}.
\end{equation}
Note that $\hat{k}\cdot\vec{S}=S_y$ is the spin-1 helicity operator and the edge photon carries a discrete unit of bosonic charge $Q_1=1$. Likewise, the electronic edge wave carries a discrete unit of fermionic charge $Q_{\sfrac{1}{2}}=\sfrac{1}{2}$,
\begin{equation}
(\sigma_y/2)\psi_\pm^\textrm{e}=\pm Q_{\sfrac{1}{2}}\psi_\pm^\textrm{e},
\end{equation}
where $\hat{k}\cdot\vec{\sigma}/2=\sigma_y/2$ is the spin-\sfrac{1}{2} helicity operator. For spin-1, helical quantization means the field is completely transverse to the momentum $\hat{k}\cdot\vec{\Psi}_\pm^\textrm{e}=0$ and the edge state behaves identically to a massless photon. This is known as the quantum gyroelectric effect (QGEE) \cite{VanMechelen2018}. Similarly, the edge electron behaves just like a helical Weyl fermion. Their dispersion relations read,
\begin{equation}
\omega_\pm(k_y)=E_\pm (k_y)=\pm vk_y, \qquad -k_i<k_y<k_i.
\end{equation}
No solution exists for $k_y\to -k_y$ and the edge states are back-scatter immune. Notice they are linearly dispersing (massless) such that the group velocity is constant $\partial_{k_y}\omega_\pm=\partial_{k_y}E_\pm=\pm v$. Moreover, the edge states are gapless and touch the bulk bands precisely at the band inversion point $k_i=\sqrt{\Lambda_0/\Lambda_2}$, where $\Lambda(k_i)=0$. At this particular momentum, one of the decay lengths becomes infinite $1/\eta(k_i)\to \infty$ and the edge states join the continuum of bulk bands. A diagram of the bulk and edge dispersion is shown in Fig.~\ref{fig:dispersion}.

\textbf{Note}: It should be pointed out that the photonic edge states $\vec{\Psi}_\pm^\textrm{e}$ are ill-defined in the zero energy limit $\omega=k_y=0$, which is characteristic of all transverse waves. This is where the edge dispersion intersects the longitudinal band $\omega_0=0$. Since this state is removed from the spectrum (no static charges present), the electromagnetic field vanishes at this point. No zero modes exist for the photon. On the other hand, the electronic edge states $\psi_\pm^\textrm{e}$ have a smooth limit at $E=k_y=0$ and zero modes are permitted. This is yet another significant difference between bosons and fermions which is related to the fact that the Dirac equation can host Majorana bound modes \cite{Oreg2010}. Since photons are their own antiparticles, no such Majorana states are possible.

\section{Conclusion}

In conclusion, we have introduced the concept of a photonic Dirac monopole appropriate for the field of spin photonics, topological photonic crystals and metamaterials. It shows magnetic monopole charge quantization in momentum space arising solely from spin-1 properties of the photon. We elucidated this phenomenon using a Dirac-Maxwell correspondence in the Riemann-Silberstein basis and applied this topological theory to 2D photonic materials. These topologically massive photons are interpreted as spin-1 skyrmions and arise from nonlocal Hall conductivity. Our work illuminates the role of photon spin in the recently proposed quantum gyroelectric phase of matter and topological bosonic phases \cite{van_mechelen_2017,VanMechelen2018}. The edge states of such a topological phase exhibit spin-1 quantization as opposed to spin-\sfrac{1}{2} quantization in fermionic phases of matter. This is ultimately connected to the presence of quantized monopole charges (bosonic- or fermionic-like \cite{Metlitski2013,Vishwanath2013}) in the dispersion of bulk matter. Experimentally probing monopole charge in momentum space can shed light on fundamental symmetries in topological electrodynamics of photons and electrons.

\section*{Acknowledgements} This research was supported by the Defense Advanced Research Projects Agency (DARPA) Nascent Light-Matter Interactions (NLM) Program and the National Science Foundation (NSF) [Grant No. EFMA-1641101].

\bibliography{dirac_monopole.bib}

\end{document}